# Accelerating Cathode Material Discovery through *Ab Initio* Random Structure Searching

Bonan Zhu[1,3,†,*], Ziheng Lu[2,3,†,*], Chris J. Pickard[2,4], David O. Scanlon[1,3,5,6]

**AFFILIATIONS**

[1] Department of Chemistry, University College London, 20 Gordon Street, London WC1H 0AJ, United Kingdom

[2] Department of Materials Science & Metallurgy, University of Cambridge, 27 Charles Babbage Road, Cambridge, CB3 0FS, United Kingdom

[3] The Faraday Institution, Quad One, Becquerel Avenue, Harwell Campus, Didcot, OX11 0RA, United Kingdom

[4] Advanced Institute for Materials Research, Tohoku University, Sendai, Japan

[5] Thomas Young Centre, University College London, Gower Street, London WC1E 6BT, United Kingdom

[6] Diamond Light Source Ltd., Diamond House, Harwell Science and Innovation Campus, Didcot, Oxfordshire OX11 0DE, UK

[†] These authors contributed equally to this work.

[*] Author to whom correspondence should be addressed bonan.zhu@ucl.ac.uk zl462@cam.ac.uk

## Abstract
The choice of cathode material in Li-ion batteries (LIBs) underpins their overall performance. Discovering new cathode materials is a slow process, and all major commercial cathode materials are still based on those identified in the 1990s. Materials discovery using high-throughput calculations has attracted great research interest, however, reliance on databases of existing materials begs the question of whether these approaches are applicable for finding truly novel materials. In this work, we demonstrate that *ab-initio* random structure searching (AIRSS), a first-principles structure prediction methods that does not rely on any pre-existing data, can locate low energy structures of complex cathode materials efficiently based only on chemical composition. We use AIRSS to explore three Fe-containing polyanion compounds as low-cost cathodes. Using known quaternary $LiFePO_4$ and quinary $LiFeSO_4F$ cathodes as examples, we easily reproduce the known polymorphs, in addition to predicting other, hitherto unknown, low energy polymorphs, and even finding a new polymorph of $LiFeSO_4F$ which is more stable than the known ones. We then explore the phase space for Fe-containing fluoroxalates, predicting a range of redox-active phases that have yet to be experimentally synthesized, demonstrating the suitability of AIRSS as a tool for accelerating the discovery of novel cathode materials.

## Introduction
The availability of reliable, safe, and accessible energy storage solutions is crucial for a world transitioning towards renewable energy sources from fossil fuels. Li-ion ion batteries (LIBs), initially developed in the 1990s, have become the leading answer to these challenges. While the manufacturing costs of LIBs have been reducing over the years, the reliance on transition metals that are expensive and sensitive to supply chain issues must be overcome for large-scale applications, such as electric vehicles (EV) and grid-scale storage. Developing the next generation cathode material to meet these requirements is a multifaceted challenge[1,2], which involves both understanding and optimising existing materials as well as discovering new ones. The three main classes of commercialised cathode materials, layered $LiCoO_2$[3], spinel $LiMn_2O_4$[4], and olivine $LiFePO_4$[5] were all proposed three decades ago, highlighting the slow nature of cathode material identification and research. Current high-performance lithium cathode materials under intensive research include: Li-rich NMC[6], disordered rock salts[7], and $\varepsilon\text{-}VOPO_4$[8].

The explosive increase in computational power in the last two decades has made computational materials research a valuable tool[9]. First-principles calculations can give invaluable insights for various properties of cathode materials[10], such as the voltages[11,12], Li-diffusion barriers[13], disordering[7], and anionic redox[14]. However, the crystal structure of the material of the interest must be known in the first place. The combination of standardised density functional theory (DFT) calculation routines and online databases have made it possible to perform so-called "high-throughput" screening studies that aim to locate new materials based on data mining of hundreds and thousands of known materials[15–17]. On the other hand, it is an open question whether such an approach can find truly "new" materials beyond those generated by simple elemental substitution. In the meantime, in other fields, such as high-pressure research where experimental data is not as plentiful, first-principles calculations have become an invaluable tool for predicting the crystal structure of unknown materials with little or no experimental data[18]. To predict stable crystal structures, the global minimum in a high dimensional potential energy surface (PES) must be located. Various algorithms have been developed for tackling this problem by exploiting intrinsic features of the PES. These methods include basin hopping[19], minima hopping[20], genetic algorithms[21], particle swarm optimisation[22] and random structure searching[23,24]. Several new data-driven approaches involving machine learning techniques have also been proposed recently, either attempting to learn the configuration spaces for generating sensible candidate structures[25,26], or use machine learning potentials as surrogates for otherwise expensive first-principles calculations[27–29].

In this article, we use several old and new cathode materials: $LiFePO_4$, $LiFeSO_4F$ and Fe-containing oxalates as examples to show that *ab initio* random structure search (AIRSS)[23,24] is a simple yet efficient tool for exploring the configuration space of complex materials and predicting the structure of existing and novel complex cathodes. We focus on Fe in this work as it is non-toxic, very abundant and low-cost, making Fe-based LIBs viable for EV and grid-scale energy storage applications. Details of the working principles of AIRSS can be found in the literature[24]. In short, this method generates random but physically sensible candidate structures, followed by geometry optimisations using density functional theory calculations. This process is repeated until the collection of low energy structures has been repeatedly found. Compared to other structure search approaches, the simplicity and the lack of iterative improvement processes ensures that AIRSS is highly parallel. While first-principles calculations are usually thought to be expensive, the actual computational cost used for search can be significantly reduced thanks to crystal symmetry being kept during the geometry optimisation, and the use of less strict convergence settings. As a result, a short time-to-result can be achieved.

To illustrate the utility of AIRSS in the field of cathode simulations, we first show that the two known experimental structures of quaternary $LiFePO_4$ can be located easily with AIRSS. Secondly, for quinary $LiFeSO_4F$, our search has located several experimentally observed polymorphs, together with previously unknown polymorphs that are predicted to be lower in energy than the known tavorite and triplite phases. One of these new polymorphs is predicted to possess both the high voltage found in the triplite phase and the three-dimensional Li diffusion network featured by the tavorite phase. Finally, we demonstrate how AIRSS can be successfully applied to determine new cathode materials with an analysis of fluorinated novel Li-stuffed Fe oxalates that have improved specific capacity and stability compared to the existing material $Li_2Fe(C_2O_4)_2$[30,31].

## Methods

*Ab initio* random structure searching (AIRSS)[23,24] was used for locating low energy structures of a given composition. This method involves generating random but physically sensible structures, followed by geometry optimisations using first-principles calculations. The random structures generated are constrained by species-wise minimum separations (Table S1) and include up to four randomly chosen symmetry operations. The polyanions are introduced as rigid units when generating the structures, but they are allowed to relax in subsequent geometry optimisations. The plane-wave density functional theory code CASTEP[3] was used for geometry optimisations. During the search, core-corrected ultrasoft pseudopotentials[32,33] are on-the-fly generated as defined in the built-in QC5 library. A plane wave cut-off energy of 300 eV is used, and the reciprocal space sampling is performed using Monkhorst-Pack grids with a maximum spacing of 0.07 $2\pi Å^{-1}$. VASP[34–37] was used for further relaxations and property calculations with a plane wave cut-off energy 520 eV, gamma-centred Monkhorst-Pack grids with a maximum spacing of 0.04 $2\pi Å^{-1}$, with projector augmented wave[38] potential dataset PBE v.54 (potentials used are *Li_sv*, *Fe_pv*, *O*, *S*, *Mn_pv* and *C*). The *Phonopy* package is used for the calculations of phonon dispersions through the finite-displacement method[39]. First-principles calculations are performed using the PBE exchange-correlation functional[40] unless otherwise stated in the text. We also performed calculations using the PBEsol functional[41] to validate the energy rankings among different polymorphs, since it gives lattice constants closer to the experimental values. Hubbard U corrections[42] are used with $U_{eff}$ = 4.0 eV for Fe d electrons[43]. For Li diffusion analysis, *ab initio* molecular dynamics (AIMD) simulations and climbing-image nudged-elastic band[44,45] calculations were performed. The MD simulations were carried out at an elevated temperature of 1200 K to enhance the sampling of the Li distribution. The temperature was controlled by coupling the system to a Nose-Hoover thermobath[46,47]. For nudged elastic band (NEB) calculations, a single Li vacancy is created in a supercell of $\sqrt{2} \times \sqrt{2} \times 2$ for $LiFeSO_4F$, containing 128 atoms in total. This ensures the minimum distance from an atom to its periodic image is to be larger than 10 Å. The force convergence criterion was set to be 0.02 eV $Å^{-1}$. Additional electrons have been added to avoid complications arising from charge localisation in the vacancy containing cells. Since the barrier height is primarily affected by local structure, this approximation is expected to give minor effects[48]. For oxalates, the band gap values are calculated using the HSE06 hybrid functional based on the PBE+U relaxed structure. The AiiDA framework[49,50] was used for managing the calculations and preserving calculation provenance. The pymatgen[51], ase[52] and sumo[53] python packages are used for manipulating structure and general analysis. The VESTA[54] software is used for visualising crystal structures.

## Results and Discussions

### $LiFePO_4$

Cathode materials are known for their chemical complexity – they typically involve at least three elements. The increasing number of elements results in a combinatorial increase of the configuration space to be explored, making structure prediction a challenging task. Here, we show that with the application of a few physical constraints, as described in the Methods, the ground state structure of known cathode materials can be obtained efficiently using AIRSS, together with many other known or unknown polymorphs.

$LiFePO_4$ (LFP) is a well-studied cathode material that has been commericallised[5]. It was originally identified in the late 90s and is now found in LIBs for electric vehicles. It has an olivine structure with space group *Pnma*. The Fe ions are octahedrally coordinated, and $FeO_6$ octahedra are corner-sharing and arranged in a zig-zag fashion. The $PO_4$ groups share edges with the $FeO_6$ octahedra. The primitive unit cell contains four formula units, giving 28 atoms in total. Our search found both the experimentally known olivine phase[5] and the high-pressure *Cmcm* phase[55,56], shown in *Figure 1*a and *Figure 1*b respectively. *Figure 2* shows energy differences and volumes of the relaxed phases with energy differences less than 50 meV per atom as compared to the ground state olivine phase. The high-pressure *Cmcm* phase is 13 meV per atom higher in energy, with a smaller volume. Structures from the Materials Project[15] database are depicted in the plot as diamonds (MP), and those from the AIRSS search are marked by dots. All structures are assumed to have ferromagnetic (FM) spin arrangements. The low-temperature ground state of LFP has anti-ferromagnetic (AFM) spin arrangements[57]. However, the energy differences between FM and AFM spin configurations are found to be only in the order of 1-2 meV for the two experimental phases. Such small energy change would have a negligible effect on the ranking of the lower energy structures.

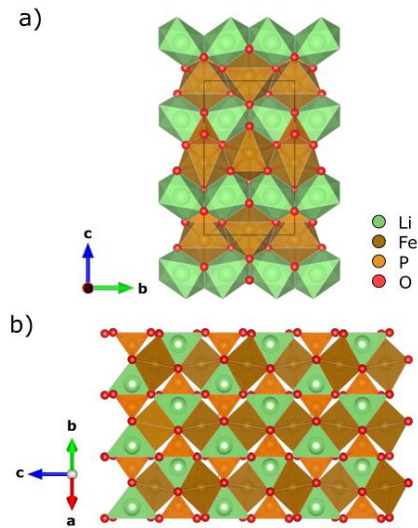

Figure 1 LiFePO$_4$ in the olivine Pnma phase (a) and the high-pressure Cmcm phase. Color-coding for atoms: Li-green, Fe-brown, P-orange, O-red.

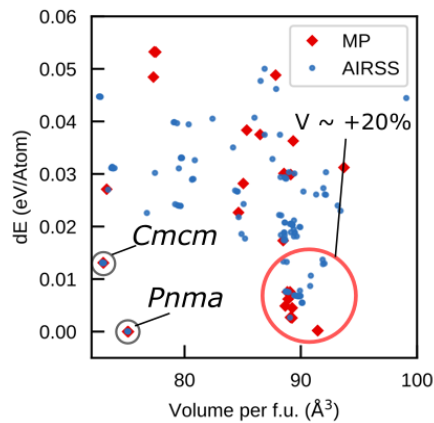

Figure 2 Energy versus volume plot for the low energy structures of LiFePO$_4$. Structures found in the Materials Project (MP) database are labelled as solid diamonds, and those found by AIRSS are represented by dots. A cluster of large volume (+20% vs Pnma) phases is also highlighted. Its apparent stability is a result of PBE exchange-correlation functional favouring less dense phases.

In addition to the two experimental phases, we also found other low energy polymorphs, and many of them are not present in the Materials Project database, as shown in *Figure 2*. To the best of our knowledge, there are no details available as to how structures in the latter were obtained, other than the experimental phases, which are based on the data from the Inorganic Crystal Structure Database (ICSD). There is a cluster of apparent low energy structures with 20% higher volume compared to the *Pnma* phase, as indicated by the red circle in *Figure 2*. These structures are characterised by networks of tetrahedrally coordinated Li and Fe atoms. Our searches also found a few polymorphs with similar features, although the apparent lowest energy phase inside this cluster is not reproduced. Care should be taken when comparing the DFT energies of candidate structures with very different densities, especially with the PBE functional which is known to systematically overpredict the lattice constants of solids[58]. After re-relaxing these phases with the PBEsol functional[41], which is designed for solids and reproduces the lattice constants better, the cluster of less dense structures becomes ~30 meV per atom higher in energy than the olivine phase (*Figure S1*). This makes them less interesting as candidates for further investigations. We chose to use PBE initially as it was more widely used in the literature, but our results here show PBEsol is a better choice for future studies involving ranking different polymorphs. Zhu et al. have previously reported finding the olivine phase using a structure prediction approach that involved motif based random structure generation with symmetry[59], which is similar to our approach here. Rather than performing DFT geometry optimisation, they included an additional step of constructing embedded atom model(EAM) potentials and use them for pre-screening, followed by final DFT relaxation on selected phases. While the fitted EAM potentials can accelerate the geometry optimisation, it is not clear whether they can faithfully reproduce features of the actual potential energy surface, which is essentially for biasing the search towards finding realistic low energy structures.

## LiFeSO$_4$F

LiFeSO$_4$F is a high voltage iron-based polyanion cathode material giving similar capacity as compared to LiFePO$_4$ but having the advantage of being more ionically and electronically conductive[60], which may remove the need for resorting to nanosizing and conductive coating of particles as in LiFePO$_4$ cathodes[60–63]. Recham et al.[60] successfully synthesised the tavorite phase using FeSO$_4$·H$_2$O and LiF as the precursors with an ionothermal method that operates at relatively low temperatures. They found the tavorite phase to be highly reversible with a voltage of 3.6 V against Li metal. Shortly afterwards, a triplite phase of the same composition was reported to have a record-high voltage of 3.9 V for Fe based polyanion cathodes[64,65]. Unlike the tavorite phase, which has well-defined Li and Fe sites[66], the triplite phase has Li-Fe site occupancy disorder, giving an entropic stabilisation. This is consistent with it being synthesised by annealing the tavorite phase[64], and later it was shown that spark-plasmas synthesis and a high-temperature solid-state route can produce it directly from the precursors[67,68]. Other polymorphs also exist for this class of material taking the general formula Li/NaMSO$_4$F (M=Co, Ni, Mn)[69–71]. The large difference in operating voltage between the tavorite and triplite phase has attracted the attention of theoretical studies[12,72–74]. Chung et al.[73] performed first-principle calculations and showed that the difference in voltage arises from the stabilities of the delithiated phases, while the lithiated structures are very similar energetically. The work of Yahia et al.[74] pointed out that the high voltage of the triplite phase can be related to the *cis* arrangement of F$^-$ ions, giving larger repulsion than the *trans* arrangement found in the tavorite phase. As the lithiated triplite and tavorite structures have similar stability, the strong F-F repulsion in the former is compensated by attractive Li-F interactions, which result in less stable delithiated phase. The secondary inductive effect caused by neighbouring Li atoms close to the O anions has been identified as an effective indicator of open circuit voltages among different Fe-based polyanionic cathodes[12]. The increased voltage of the triplite phase is attributed to having an increased number of Li neighbours around the Fe-centred octahedra when compared to that of the tavorite phase.

The rich set of structure-property relationships in polymorphs of LiFeSO$_4$F leads one to wonder if there are yet more phases to be discovered, potentially with even better electrochemical properties. The involvement of F further increases the complexity of the search, making it a quinary system. The two known experimental phases, tavorite and triplite, are displayed in Figure *3a* and Figure *3b* respectively. The relative energies and volume per formula unit of the structures found in the search are shown in Figure *3c*. To mitigate the issue of PBE favouring less dense structures, as it was found in the search of LiFePO$_4$ above, we searched for ambient pressure as well as with a 10 GPa external pressure. There are two experimental tavorite structures with slightly different Li atom arrangements. The initial report of the tavorite phase proposed that the Li atoms occupy two half-occupied *2i* Wycoff sites[60], but later neutron diffraction data show they occupy a single *2i* site[66]. This structure is labelled as Ta-I, and it has lower energy than the former in an ordered unit cell (Ta-II). Our search found two more variants that have slightly different Li sites but similar energies (phase F/F'). The triplite phase (space group *C2/m*) has been reported to exhibit Li-Fe site-occupation disordering[64]. To obtain a meaningful theoretical model, unique Li-Fe configurations in the primitive cell are enumerated by distributing four Fe and four Li atoms. These structures are labelled as green crosses in Figure *3c*. The lowest energy structure is shown in Figure *3a* (Tr-I), which has a cation arrangement consistent with the work by Chung et al.[73]. The same structure is also obtained in the search.

New low energy polymorphs have been found by our computational search, and many of them have comparable or even lower energy compared to the experimental phases, as shown in Figure *3c*. Structures A-D have FeO$_4$F$_2$ octahedra arranged in a configuration containing one-dimensional edge-sharing chains. Phase A (*Figure 4a*) has the lowest energy. It has F ions arranged in a *trans* fashion in the FeO$_4$F$_2$ chains, which are separated by isolated SO$_4$ groups and tetrahedral coordinated Li atoms. The flexibility in the octahedra chains and separators result in different polymorphs with similar energies. The B phase (*Figure 4b*) has similar features but the FeO$_4$F$_2$ chains are arranged in a different pattern. Both have larger volumes compared to the known tavorite/triplite phases. On the other hand, the C phase (*Figure 5a&b*) is more efficiently packed. In fact, it closely resembles the sillimanite Al$_2$SiO$_5$ structure[75]. A variant of this structure, phase D (*Figure 5c*), has a different occupation of Li and S sites between the FeO$_4$F$_2$ chains. Previously, LiZn$_x$Fe$_{1-x}$SO$_4$ has been found to crystallise in the sillimanite structure, where x can be as high as 0.15 using solid-state synthesis, or 0.1 using an ionothermal approach, before the tavorite/triplite phase starts to crystallise instead[69]. The reported structure of this phase has Li atoms in the octahedral site and Fe atoms in the tetrahedral sites, e.g., the opposite of phase C&D. To avoid confusion, from now on we refer the sillimanite structure found here as sillimanite-II, and the previously reported LiZn$_x$Fe$_{1-x}$SO$_4$ phase as sillimanite-I. The latter is also rediscovered in our search and labelled as structure E (*Figure 5e*).

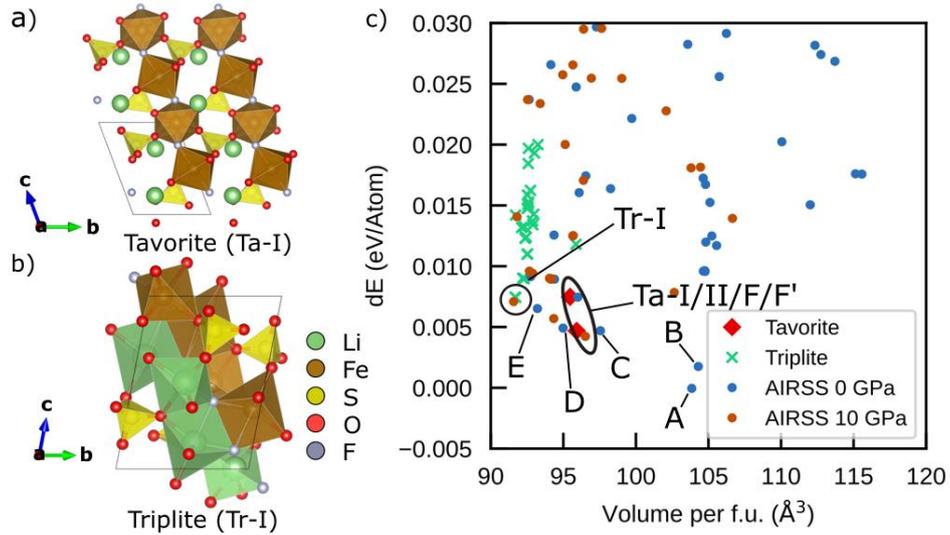

Figure 3 The two known experimental phases of LiFeSO$_4$F, tavorite (a) and triplite (b), are both found in the search. Relative energy per atom is plotted against the volume per formula unit in (c). In addition to the experimentally known phases, several new polymorphs are found using AIRSS. Strucutres found by searching at 10 GPa are futher relaxed at the ambient pressure. Colour-coding for atoms: Li-gree, Fe-brown, S-yellow, O-red, F-grey.

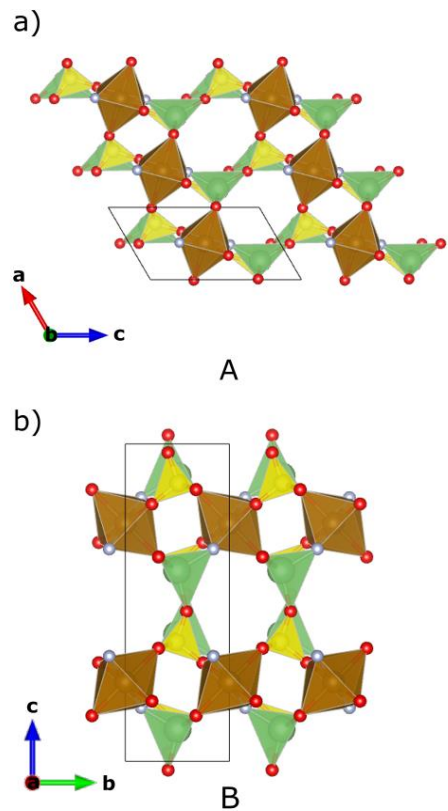

Figure 4 Phase A (a) and B (b) found using AIRSS. Both have edge-sharing chains of FeO$_4$F$_2$ along the c direction, but the chains are arranged differently relative to their neighbours. Color-coding for atoms: Li-green, Fe-brown, S-yellow, O-red, F-grey.

A factor that was neglected in the initial search and the analysis so far is the spin configuration, as all sampled structures are assumed to be ferromagnetic. To assess the effect of spins, collinear magnetic configurations have been enumerated to find the low energy AFM states using the method implemented in the *pymatgen* package[76]. The energy differences (with respect to the FM state of phase A) for both AFM and FM state is shown in *Figure S2*. The energy difference between the AFM and FM spin state for each phase is tabulated in *Table 1*. Changing FM to AFM spin arrangement reduces the total energy, with the magnitude of the reduction typically about 1 meV, which is inconsistent with those found for LiFePO$_4$ above. Although in some cases, it can be up to 5 meV per atom. This can be understood as the potential energies are mainly contributed by the electrostatic energies and short-range repulsions in ionic crystals. As a result, assuming FM ordering in the initial search would have a relatively small effect on the overall ranking of the structures.

Since phases A-D all share similar features but differ in their volumes by up to 10%, it raises the question of whether the seemingly better stability of phase A, which has a larger volume, is an artefact of the PBE exchange-correlation functional. This was found to be the case for the polymorphs of LiFePO$_4$ as discussed above. Additional relaxations have been performed using the PBEsol functional, keeping the same FM spin configurations. The energy difference between phase A and C is now reduced by ~ 5 meV per atom, and the latter becomes the lowest energy structure, as shown in *Figure S3*. The change in energy here is much smaller than those for LiFePO$_4$ polymorphs. Unlike LiFePO$_4$, despite subtle changes in the rankings, the collection of low energy structures remains the same as those using the PBE functional. Further checks using hybrid functional HSE06 shows that phase C has lower energy compared to the known sillimanite-I structure (phase E), as well as the tavorite and triplite structures (*Figure S4*).

Selected properties of the different polymorphs are tabulated in *Table 1*. The average voltages of the tavorite phases are found to be 3.52 eV, which is in good agreement with other theoretical studies[72,73] and the reported value of 3.6 V measured from experiments[60]. We note that the voltages calculated are positively correlated to the choice of the U parameter in the GGA+U. The FM spin configuration is used since the energy differences in AFM-FM spin configurations give an insignificant effect on the voltage. The initial atomic positions of the delithiated cells are shaken by an amplitude of 0.05 Å to break the symmetry. For the triplite phases, we found their voltages to range from 3.9 V to 4.3 V depending on the Li-Fe orderings. The most stable Fe orderings are likely to be different in the lithiated and delithiated cells. These values are in consistent with the experimental reported 3.9 V[64]. For structure E (sillimanite-I), the average voltage is found to be 3.51 V, which is in good agreement with the experimental value of 3.6 V[69]. On the other hand, those in the sillimanite-II structure, phase C and D, have a higher voltage of 4.0 V. Their volumes are reduced by 5.7% and 4.9% respectively after delithiation, which are lower than that of the tavorite phase (Ta-I). On the other hand, the triplite structures have much smaller volume changes upon delithiation, which can be related to Li-Fe disordering and being more densely packed in the first place. The positive volume changes displayed in *Table 1* is likely to be a result of only considering removing Li in specific Li-Fe orderings, hence they are not truly representative of a truly disordered system.

*Table 1 Properties of the LiFeSO$_4$F polymorphs found in the search and known previously. ΔE$_{FM}$, ΔE$_{AFM-FM}$, ρ, V$_{avg}$ and ΔVol are energy differences to the lowest energy phase (phase A) assuming FM ordering, the energy difference between FM and AFM spin configuration, density, average voltage and volume change in the delithiated phases respectively. Phase A-D contain a FeO$_4$F$_2$ chain with C&D in the sillimanite-II structure. Phase E is identical to the previously reported sillimanite-I structure for LiFe$_{1-x}$Zn$_x$SO$_4$F. Polymorphs labelled as Tr-I/II/III are triplite structures with different Li-Fe orderings. Phase Ta-I, Ta-II and F all have the tavorite framework, but Li atoms occupy slightly different sites. Phase Ta-I is the reported tavorite structure with a single Li Wyckoff site.*

| Polymorph | ΔE$_{FM}$ (meV) | ΔE$_{AFM-FM}$ (meV) | ρ (g cm$^{-3}$) | V$_{avg}$ (V) | ΔVol (%) |
|---|---|---|---|---|---|
| A | 0.0 | -0.2 | 2.84 | 4.01 | -1.84 |
| B | 1.7 | -0.0 | 2.83 | 4.00 | -3.2 |
| Ta-I | 4.7 | -1.3 | 3.09 | 3.52 | -7.4 |
| C | 4.8 | -4.7 | 3.03 | 4.02 | -5.7 |
| D | 4.9 | -0.4 | 3.11 | 4.05 | -4.9 |
| E | 6.5 | -1.1 | 3.17 | 3.51 | -5.6 |
| Tr-I | 7.5 | -0.2 | 3.22 | 4.21 | 4.0 |
| F | 7.5 | -1.4 | 3.08 | 3.50 | -6.2 |
| Ta-II | 7.5 | -1.4 | 3.10 | 3.50 | -6.7 |
| Tr-II | 9.0 | -0.4 | 3.20 | 4.13 | 0.18 |
| Tr-III | 11.0 | -0.0 | 3.19 | 3.88 | 6.58 |

We now turn our focus to the sillimanite-II structured C phase which has edge-sharing chains of $FeO_4F_2$ octahedra and separated by chains of corner-sharing $SO_4$ and $LiO_3F$ tetrahedra. This phase has space group $P2_1/c$ - a subgroup of *Pnma* in the original sillimanite $Al_2SiO_5$ phase. Its symmetry is further lowered to $P\bar{1}$ when Fe atoms are in an anti-ferromagnetic spin arrangement. Finite-displacement phonon calculations show it is dynamically stable since no imaginary frequencies are found across the first Brillouin zone, as shown in *Figure S5*. Compared to the previously reported sillimanite-I structure (phase E), the sillimanite-II structured phase C has a higher voltage (4.0 V vs 3.5 V). The increased voltage means a higher energy density for cathode applications.

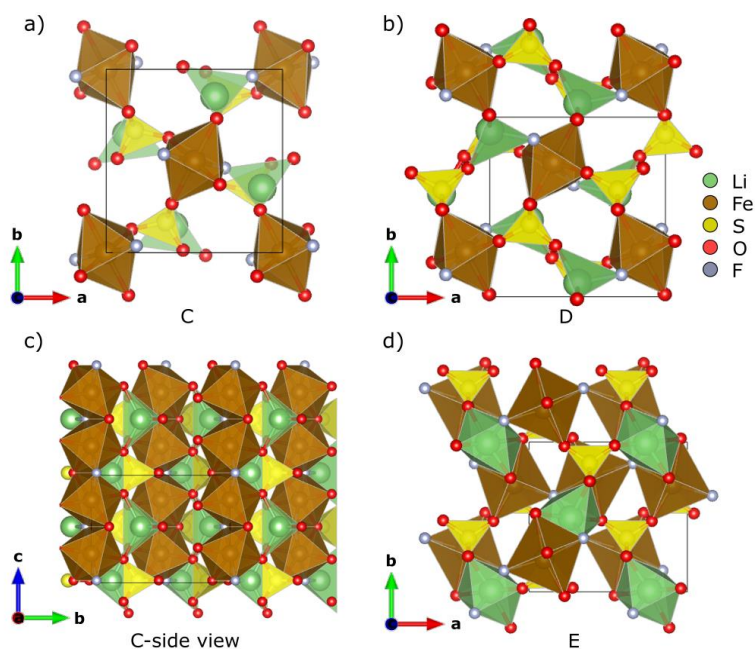

Figure 5 Phase C when viewed along the **c** direction (a) and the **a** direction (c), which resembles the sillimanite $Al_2SiO_5$. A closely related structure is phase D (b), it has a different Li and S arrangement among the tetrahedral sites. Phase E (d) is the fully occupied version of the "sillimanite" $LiFe_{1-x}Zn_xSO_4F$, where the occupation of Li and Fe is the reverse of phase D. Color-coding for atoms: Li-green, Fe-brown, S-yellow, O-red, F-grey.

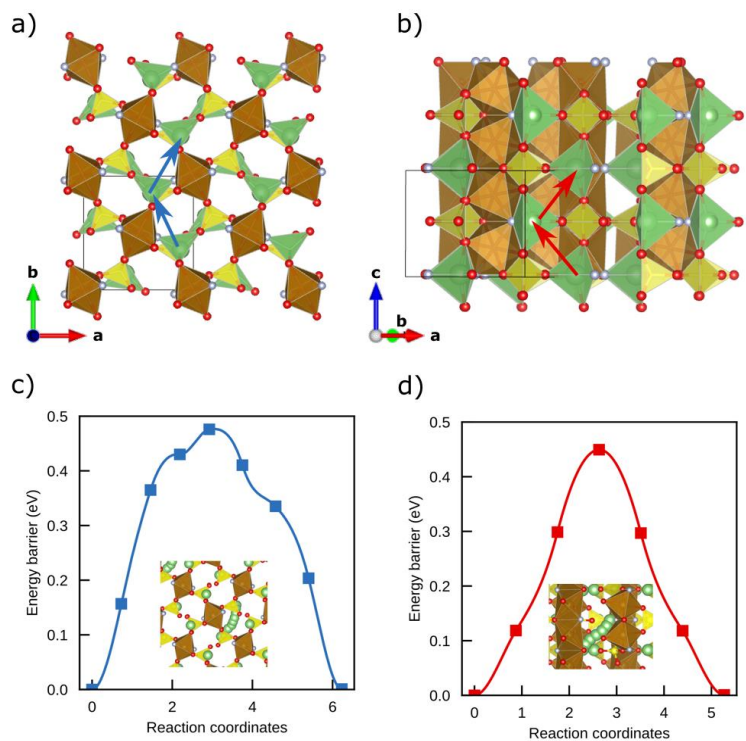

Figure 6 Li diffusion pathways and barrier heights in the sillimanite-II structured LiFeSO$_4$F (phase C). Colour coding: green-Li, red-oxygen, brown-Fe, yellow-S, grey-F.

While the triplite phase also has a relatively high voltage, Li-Fe disordering limits the reversible extraction and insertion of Li, giving a lower reversible Li content as compared to the tavorite and the sillimanite-I phase[60,64,69]. In the sillimanite-I structure, the nearest neighbour Li atoms are along the c direction, and the second nearest neighbour Li atoms are much further apart as compared to the nearest neighbours (3.2 Å vs 5.2 Å). This implies Li atoms are constrained to diffuse along 1D channels. In contrast, the first and second nearest neighbours of Li atoms are 4.3 Å and 4.5 Å away in sillimanite-II structured phase C. We have identified two potential Li-vacancy hopping pathways, shown in Figure *6a* and Figure *6b* respectively. Path A involves the zig-zag movement of Li atoms along the [010] direction. To hop to the next site using a direct path, the Li atom must go through an edge and a face of the LiO$_3$F tetrahedra. The NEB calculations indicate that the transition state barrier height of this process is 0.48 eV, as shown in Figure *6c*. The shape of the curve is a result of the Li atom taking an indirect route – it hops around the edge of the coordination tetrahedron rather than squeezing through it. The results of NEB calculations for path B are shown in Figure *6d*. In this case, the overall mass transport takes place along the [001] direction, although individual Li atoms also take zig-zap paths. The barrier of a single vacancy hop is found to be 0.45 eV and the symmetrical shape of the curve is a result of the initial and final images being related by the inversion symmetry. Because a single hop along path B would take the Li atom to the next chain of path A in the [101] direction, it is also possible to have net Li diffusion along the [100] direction. We also performed AIMD simulations and found that the distribution of Li forms a zig-zag 3D network (*Figure S6*) in good agreement with the two NEB paths identified. The relatively low average barrier height, 0.45 eV, make phase C a promising material, with good ionic conductivity. In comparison, the barrier in LiFePO$_4$ was found to be about 0.55 eV[77], the tavorite and triplite LiFeSO$_4$F was reported to have barriers of 0.57 eV and 0.6-0.8 eV respectively[78], and that of Li$_2$FeSiO$_4$ was reported to be 0.91 eV[79]. We note that the exact values of the barrier height can be affected by the details of the methods used, i.e., the level of theory and treatment of vacancy defects.

The relatively small energy differences between sillimanite-I and sillimanite-II phases raise the question as to whether it is possible to have Li-Fe occupancy disordering. Enumeration of all symmetrically unique Li-Fe arrangements phase C-E show deviations from the original ordering would result in energy change of at least 0.15 to 0.2 eV per primitive cell (~5 meV per atom), which is higher than that of the triplite phases (1.5 meV per atom). The results are shown in *Figure S7*. While detailed studies of Li-Fe anti-sites in the sillimanite-II structure are beyond the scope of this work, such defects should not significantly impact the Li diffusivity due to the three-dimensional nature of the diffusion pathways.

Our search has uncovered a rich landscape of low energy polymorphs for LiFeSO$_4$F, and a new sillimanite-II phase is predicted to have both the high voltage of the triplite, and fast Li-diffusion kinetics of the tavorite. This leads to the question: why has this phase not been synthesised yet? Solid-state synthesis is known to be limited by the slow reaction rate and limited mass transport. The most widely reported synthesis routes for LiFeSO$_4$F starts from FeSO$_4$·H$_2$O and LiF, where the former is already in the tavorite structure, making formation of tavorite LiFeSO$_4$F favoured via topotactical replacement of O$^{2-}$ in the H$_2$O by F$^-$ in conjunction with the intercalation of Li$^+$ ions[60]. The triplite phase is generally synthesised from the tavorite LiFeSO$_4$F by annealing or directly from the anhydrous FeSO$_4$ and LiF using spark-plasma synthesis or ball milling[67]. This can be rationalised by the entropic contributions from the Li-Fe disordering making the triplite phase favoured at high temperature. In addition, the sillimanite-I phase also has structural similarities to the tavorite phase, as it also has the corner-sharing Fe(Zn)O$_4$F$_2$ octahedra[69]. The sillimanite-II phase, on the other hand, contains the distinct feature of edge-sharing FeO$_4$F$_2$ chains not found in the other polymorphs, and the small energy difference compared with the tavorite makes the thermodynamic driving force relatively low for its formation. Nevertheless, we note that the sillimanite-II phase is in fact structurally closely related to the anhydrous FeSO$_4$. Future synthesis works may attempt to exploit this similarity, although it will still be challenging to avoid forming the triplite phase if high temperature synthesis is required.

## Transition metal oxalates

The commercial success of lithium-iron-phosphate cathodes has triggered extensive research interest in searching for polyanion compounds as cathode materials for batteries due to their low cost, long cyclic life, and high safety[80]. To date, a wide spectrum of polyanions have been intensively studied, including phosphate (PO$_4$)$^{3-}$, sulfate (SO$_4$)$^{2-}$, and silicate (SiO$_4$)$^{2-}$ [81–84]. Despite the significant effort, another family of polyanion compounds have been largely overlooked, *i.e.*, the oxalates (C$_2$O$_4$)$^{2-}$. In fact, to the best of our knowledge, only two iron-based oxalates, *i.e.*, Fe$_2$(C$_2$O$_4$)$_3$·4H$_2$O[85] and Li$_2$Fe(C$_2$O4)$_2$[30] have been reported as cathodes for LIBs while the electrochemical performance of other transition metal oxalates is largely missing. Interestingly, if one considers the polarizability of the oxalate group, it is comparable to that of (PO$_4$)$^{3-}$. Therefore, these anions should exert a strong inductive effect on transition metals thereby providing competitive redox potentials during battery discharge.

The oxalates are also characterized by a few other advantageous features when used as cathodes for batteries. One of the most obvious is the ease of synthesis. Due to the varied solubility of oxalic acid and oxalate metal salts, oxalate-based compounds can be synthesized under relatively mild conditions (usually below 200°C) via solution-based methods such as hydrothermal synthesis. This exerts advantages over conventional inorganic polyanion compounds such as LiFePO$_4$ which requires solid-state sintering at elevated temperatures of up to 700°C. In fact, it is well-known in synthetic and structural chemistry the oxalates can crystallize into a wide variety of different polymorphs by serving as monodentate, bidentate, tridentate, or tetradentate ligands. By complexing with different metal centres, the oxalates could form one-dimensional chain-like, two-dimensional layered, and three-dimensional connected frameworks[30]. Such polymorphism provides yet another exciting opportunity to tune the structural stability and the electrochemical performance of the material. In particular, such structural features have a profound influence on cation mobility and can be used to enhance the rate capability of the cathode. For example, in a recent contribution, it is shown that by reducing the dimensionality of the host framework, superionicity can be achieved in lithium-rich anti-perovskites[86–89]. Beyond structural considerations, the polyanionic nature and the covalent characteristics of the C-O bonds in oxalate anions offer exciting opportunities to tune the electrochemistry as well. Recently, anionic redox (mostly oxygen) has been discovered as an important source of additional capacity beyond the transition-metal contribution during battery discharge[90]. Unfortunately, in conventional oxide materials such as $Li(Li_xNi_{1-x-y-z}Mn_yCo_z)O_2$ (Li-rich NMC), the stripping of electrons from the oxygen non-bonding states leads to cyclic instability and sometimes oxygen release. In this regard, polyanionic species which bear covalently bonded oxygen, in principle, should have better anti-$O_2$ release capability and better reversibility in anionic redox.

All these features make oxalates a valuable family of cathode materials to explore. Nevertheless, they are less studied as a cathode material and only started to gain attention very recently. Ahouari et al.[85] first looked at the redox and structural evolution of a commercially viable Fe (III) oxalate, e.g. $Fe_2(C_2O_4)_3 \cdot 4H_2O$, during lithiation and showed that these compounds demonstrate an average voltage of 3.35V vs. Li/Li$^+$. However, due to the high concentration of coordinated $H_2O$ in such materials, the gravimetric capacity is relatively low and only reaches 98 mAh g-1. Moreover, the study was restricted to relatively narrow voltage cut-offs and the potential capacity contribution from oxygen was not looked at. Very recently, Jiao and co-workers synthesized a family of Li-stuffed Fe (II) oxalates with a composition of $Li_2Fe(C_2O_4)_2$[30]. Interestingly, they show that more than one Li can be taken out of the structure reversibly and anionic redox is incorporated during the process. We recently carried out a thorough search of polymorphs on the composition of $Li_2Fe(C_2O_4)_2$ and found several relatively stable phases that are potentially synthesizable[31]. Despite these pioneering efforts, oxalate cathodes are still in their infancy. Many more compositions are yet to be looked at. The additional chemical and configuration space which may yield possible polymorphs with excellent electrochemical performances have never been explored as well. In this contribution, we take advantage of the efficient AIRSS algorithm and conduct searches with two compositions that have not been reported before: $LiFeC_2O_4F$ and $Li_2FeC_2O_4F_2$. The main motivation for incorporating with F$^-$, other than expanding the chemical space, is to increase the theoretical capacity, since F$^-$ has a lower mass-to-charge ratio (19 e$^-$/u) compared with $C_2O_4$ (44 e$^-$/u). Fluoroxalate materials have been proposed for cathodes of potassium-ion batteries and found to have exceptional cyclability[91]. By placing a special focus on the Fe (II) containing compounds, we look for cheap, high-rate, easy-to-synthesize, and reasonably energy-dense cathodes. Specifically, the keep the problem computational tractable, we look at both compositions with the number of formula units no larger than four. The $C_2O_4$ groups are fixed as units for the sole purpose of generating the initial random structures, which are constrained to have two to four symmetry operations.

In Figure 7, we show the lowest energy structures obtained for $Li_2FeC_2O_4F_2$ and $LiFeC_2O_4F$. $Li_2FeC_2O_4F_2$ has a theoretical specific capacity of 274 mAh/g, assuming all Li can be extracted, which is higher than that of the previous report $Li_2Fe(C_2O_4)_2$ (218 mAh/g). Both $Li_2FeC_2O_4F_2$ and $LiFeC_2O_4F$ contain infinitely extending chains formed by six-fold coordinated Fe$^{2+}$ where neighbouring octahedra are connected by the oxalate group. The Li atoms are tetrahedrally coordinated by O and F. The corner-sharing Li tetrahedra forms chains that are parallel to that made of Fe centred octahedra. The dynamic stabilities of the predicted structures are confirmed by finite-displacement phonon calculations, which have found no imaginary modes across the first Brillouin zone for both (Figure S8). Our calculation shows that the $Li_2FeC_2O_4F_2$ is metastable with a distance-to-hull of 55 meV per atom, which is slightly higher than the previously reported $Li_2Fe(C_2O_4)_2$ phase which is 51 meV per atom above the hull. The metastability of these materials can be attributed to the $(C_2O_4)^{2-}$ group. Even $Li_2C_2O_4$, which is known experimentally, is also predicted to have an energy above the convex hull (47 meV per atom). The existence of these metastable phases can be understood as breaking the C=C bond requires overcoming a large energy barrier. On the other hand, $LiFeC_2O_4F$ is more unstable with a higher distance-to-hull of

83 meV per atom. Since the reaction: LiFeC$_2$O$_4$F + LiF → Li$_2$FeC$_2$O$_4$F$_2$ has negative reaction energy, it is unlikely to be synthesised. Therefore, we focus only on Li$_2$FeC$_2$O$_4$F$_2$ from now on.

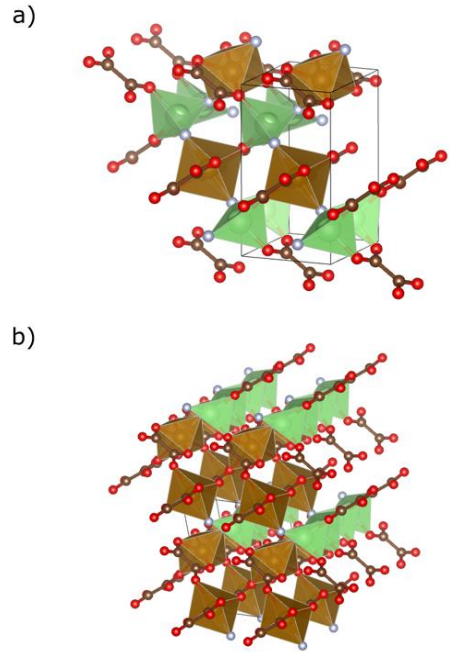

Figure 7 (a) The structure of Li$_2$FeC$_2$O$_4$F$_2$. (b) The structure of LiFeC$_2$O$_4$F. Color-coding for atoms: Li-green, Fe-brown, C-dark brown, O-red, F-grey.

A cathode material must be both electronically and ionically conductive. Using the hybrid HSE06 functional, the bandgap of Li$_2$FeC$_2$O$_4$F$_2$ is computed to be about 2.8 eV, which is lower than other polyanion materials such as LiFePO$_4$ and is comparable to materials known to have decent electronic conductivities, for example, LiCoO$_2$[92]. For ionic conduction, two Li diffusion pathways can be identified, one going along the Li-chains (Path 1 and 2) and the other across them (Path 3 and 4), as shown in Figure 8a and b respectively. Path 1 and Path 2 both have relatively low barriers of 0.39 eV with Li atoms move through faces of the coordination tetrahedron. On the other hand, Path 3 and Path 4 have much higher barrier heights (0.82 eV and 1.31 eV), which can be attributed to long hopping distances and the need for the Li atom to go through tetrahedral edges. Hence, the movement of Li is likely to be constrained inside the 1-D channels along the $a$ direction shown in Figure 8a.

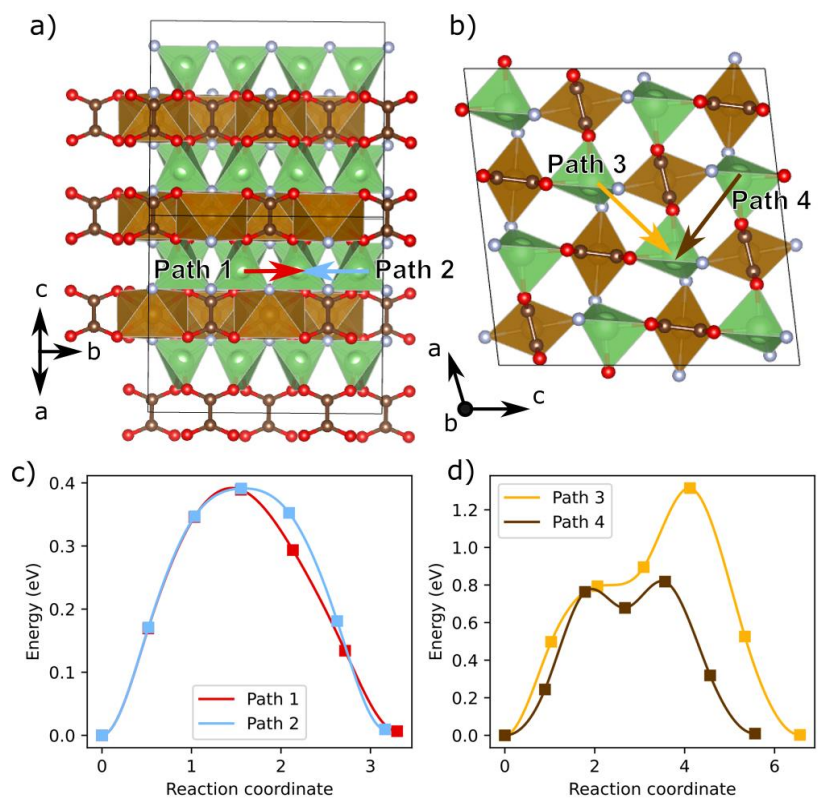

Figure 8 Potential diffusion pathways (a) along and (b) across the chains of tetrahedral coordinated Li atoms. NEB calculations show that the former has reasonable barriers heights of 0.40 eV, while the those for the latter are 0.82 eV and 1.31 eV. Color-coding for atoms: Li-green, Fe-brown, C-dark brown, O-red, F-grey.

The stabilities of delithiated phases of $Li_{2-x}FeC_2O_4F_2$ ($0 \leq x < 2$) are investigated by enumerating unique configurations of the lithiated phase with Li atoms removed in 2x1x1 and 1x2x1 supercells (four formula units). Each structure generated is subsequently relaxed following the specification outlined in the methods section. Convex hull constructions show that the only stable composition in between the terminal compositions is $LiFeC_2O_4F_2$, resulting in two plateaus in the voltage profile, each originates from a two-phase reaction (Figure 9a). The voltage of the first step ($0 < x < 1$) is found to be 3.52 V, and the volume of the delithiated structure changes by ~1%. This gives a theoretical energy density of 482 Whkg$^{-1}$, which is higher than the 425 Whkg$^{-1}$ of $Li_2Fe(C_2O_4)_2$ (TM redox only, 3.9 V). Further delithiation requires a much higher voltage of 5.3 V. In comparison, removing more than one Li from $Li_2Fe(C_2O_4)_2$ was reported to require only a minor increase of the voltage (up to ~4.2 V)[30], and the extra capacity has been attributed to anionic redox. In contrast, no evidence of anionic redox is found for $Li_2FeC_2O_4F_2$ here based on GGA+U calculations. The Fe-projected magnetisations in delithiated structures are shown in Figure *10*a. Within the range of $0 < x < 1$, the increase of $\mu_b$ is consistent with $Fe^{2+}/Fe^{3+}$ redox, and the reversal beyond $x > 1$ indicates that the $Fe^{3+}/Fe^{4+}$ redox is active. The same trend is also found for the total magnetisation of the entire cell. The inactivity of the $(C_2O_4)^{2-}$ is confirmed by tracking the C=C bond lengths, which stay almost unchanged as more Li atoms are taken away. In contrast, removing more than one Li per Fe in $Li_2Fe(C_2O_4)_2$ leads to spontaneous breaking of a $(C_2O_4)^{2-}$ group into two $CO_2$ like parts, giving rise to increased average C-C distances, as shown in Figure *10*b. The superior C=C bond stability in $Li_2FeC_2O_4F_2$ is further supported by the Fukui functions calculated for delithiated structures with $x = 1$, where the electron density of the C=C bond has a much smaller contribution compared to that in

Li$_2$Fe(C$_2$O$_4$)$_2$, shown in Figure S9. Similar results are obtained using HSE06 hybrid functional, where the (C$_2$O$_4$)$^{2-}$ groups only spontaneously break at very high delithiation levels (Figure S10).

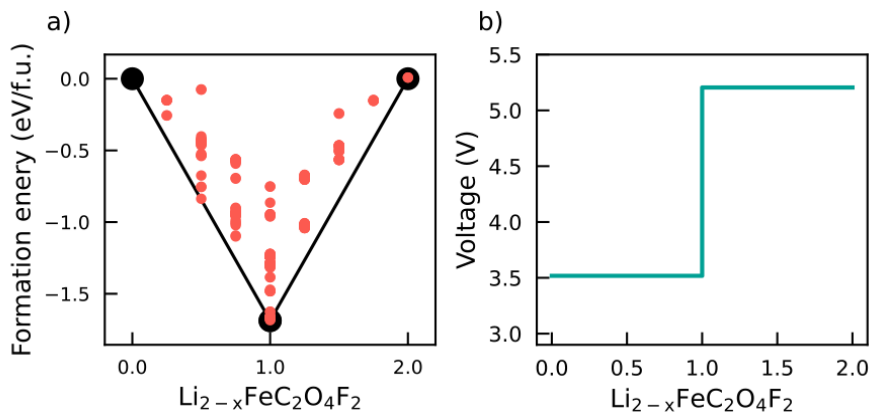

Figure 9 Calculated convex hull of the delithiated structures of Li$_{2-x}$FeC$_2$O$_4$F$_2$ (a) and the voltage profile constructed from it (b).

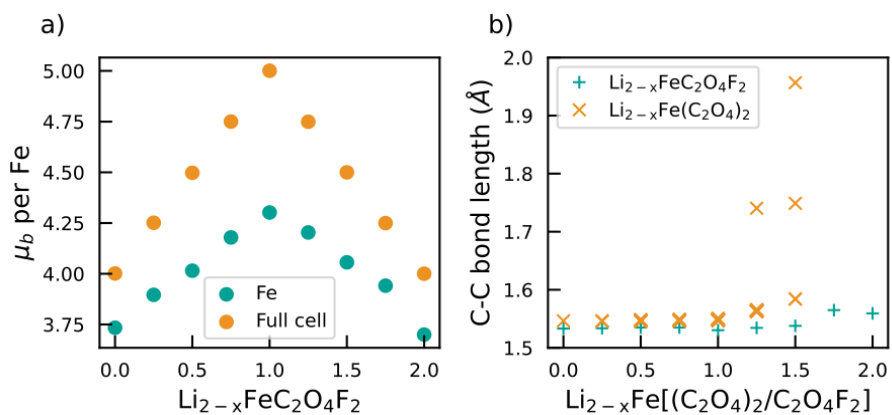

Figure 10 (a) Total magnetisation per Fe ("Full cell") and locally projected magnetisation (Fe) at different delithiation levels for Li$_{2-x}$FeC$_2$O$_4$F$_2$. (b) Bond lengths in delithiated structures for Li$_{2-x}$FeC$_2$O$_4$F$_2$ and Li$_{2-x}$Fe(C$_2$O$_4$)$_2$.

Our prediction of Li$_2$FeC$_2$O$_4$F$_2$ opens a new avenue for developing transition metal oxalate-based cathode materials and demonstrates the effectiveness of AIRSS in exploring cathode materials with new compositions. The incorporation of F$^-$ ion is shown to improve the specific capacities and stabilise the oxalates group from dissociation at high levels of delithiation. Despite no anionic redox being observed, our predictions provide a model system for future work to unravel the nature of the anionic redox in polyanions.

## Conclusions

We have shown that *ab initio* random structure searching (AIRSS) can be used to predict the low energy polymorphs of complex cathode materials efficiently. For LiFePO$_4$, the two experimental phases are rediscovered in the search, along with many other unknown low energy polymorphs. Our searches for LiFeSO$_4$F have not only rediscovered several experimental phases but also found new sillimanite structured polymorphs with a sillimanite structure with edge-sharing FeO$_4$F$_2$ chains. One of these new phases is predicted to have both a higher voltage (~4.0 V) and a 3D Li diffusion network with relatively low barrier heights, combine the advantage of both the existing tavorite and triplite phases. When applied to fluorinated Li-Fe oxalates, our search has found a $Li_2FeC_2O_4F_2$ phase with 1-D Li diffusion pathways and an average voltage of 3.5 V utilizing the $Fe^{2+/3+}$ redox couple. The incorporation of $F^-$ increases the specific capacity and mitigates the decomposition of $(C_2O_4)^{2-}$ groups at high delithiation levels, resulting in improved structural stability during cycling.

First-principles structure predictions are relatively computationally demanding. Given that eventual cubic scaling of plane wave DFT, the computational cost can increase significantly with the number of atoms in the unit cell, not to mention the fact that more structures will need to be sampled. Nevertheless, we found that systems up to 30-40 atoms can be addressed at moderate costs, and many inorganic crystalline materials have unit cells that fall within this range. While making predictions from existing databases though substitution is relatively "cost-free", such an approach is not applicable to underexplored chemical spaces. Existing structural motifs and design rules[93,94] can be utilised to choose chemical systems with the potential to be high-performance cathode materials, followed by explorative searches to find synthesisable phases in targeted chemical spaces.

Cathode materials often exhibit site-occupation disordering because of similarities in ionic radii. It is difficult to addresses disordered phases directly by DFT, and therefore finding them directly via an AIRSS search is unlikely, however, they can still be identified through the existence of an ensemble of relaxed structures with similar energies sharing common frameworks. Subsequently, methods as special quasi-random structure[95] and cluster expansion[96–99] can be applied to test if the system is truly disordered.

Our approach is not limited to Li-ion intercalation cathodes, crystal structures of Na/K containing cathodes can be searched similarly. Admittedly, finding new phases theoretically, i.e., demonstrating the existence of a low energy local minimum, does not guarantee that the phase can be realised, as kinetic barriers play a very important role in synthesis. Nevertheless, knowing the existence of possible new phases will undoubtedly help to inspire and guide future experimental works. We hope that by combining predictive computational approaches and experimental efforts, the discovery of new novel cathode materials will be greatly accelerated.

# Supplementary Material
See the supplementary materials for the details of the $LiFePO_4$ search, tabulated species-wise separations, and additional calculation results

# Data Availability Statement
The data that support the findings of this study are openly available in Zenodo at
http://doi.org/10.5281/zenodo.5572347.

# Acknowledgement

This work was supported by the Faraday Institution grant number FIRG017 and used the MICHAEL computing facilities. C. J. P. acknowledges support from the EPSRC through the UKCP consortium (EP/ P022596/1). D. O. S. acknowledges support from the European Research Council (Grant 758345). Through our membership of the UK's HEC Materials Chemistry Consortium, which is funded by EPSRC (EP/L000202, EP/R029431, EP/T022213), this work used the ARCHER UK National Supercomputing Service (http://www.archer.ac.uk), the UK Materials and Molecular Modelling Hub which is partially funded by EPSRC (EP/P020194 and EP/T022213). The authors also acknowledge the use of the UCL Myriad and Kathleen High Performance Computing Facility (Myriad@UCL, Kathleen@UCL), and associated support services, in the completion of this work.

# Supplementary information

# Accelerating Cathode Material Discovery using *ab initio* Random Structure Searching

Bonan Zhu[1,2,†,*], Ziheng Lu[2,3,†,*], Chris J. Pickard[2,3,4], David O. Scanlon[1,2,5,6]

**AFFILIATIONS**

[1] Department of Chemistry, University College London, 20 Gordon Street, London WC1H 0AJ, United Kingdom

[2] The Faraday Institution, Quad One, Becquerel Avenue, Harwell Campus, Didcot, OX11 0RA, United Kingdom

[3] Department of Materials Science & Metallurgy, University of Cambridge, 27 Charles Babbage Road, Cambridge, CB3 0FS, United Kingdom

[4] Advanced Institute for Materials Research, Tohoku University, Sendai, Japan

[5] Thomas Young Centre, University College London, Gower Street, London WC1E 6BT, United Kingdom

[6] Diamond Light Source Ltd., Diamond House, Harwell Science and Innovation Campus, Didcot, Oxfordshire OX11 0DE, UK

[†] These authors contributed equally to this work.

[*] Author to whom correspondence should be addressed: bonan.zhu@ucl.ac.uk and zl462@cam.ac.uk

## Additional details of the LiFePO$_4$ search

The search for LiFePO$_4$ ground state structure involves 684 relaxed structures including four formula units and 1112 structures that have two formula units. The random structures are generated such that they have two to four symmetry operations with rigid PO$_4$ units and satisfy a predefined minimum species-wise separation. The latter was initially inferred from the known olivine phase. However, we note that this information, unlike the distance matrix, does not encode any details of the exact structure. It is common to infer their values based on the bond lengths in known materials, in the spirit that species in chemically similar compounds should have similar local environments. For completely unknown systems, brief explorative searches can be performed with randomly chosen minimum species-wise separation. Subsequently, values from the low-energy structures can be used for further searching. We do not apply any constraints during the geometry optimisation. The olivine structure has been found 4 times out of the 684 relaxed structures with four formula units, giving an average encounter rate of

1/171. The high-pressure phase has been found 9 times out of the 1112 sampled structures with two formula units, giving an average encounter rate of 1/124.

Specie-wise minimum separations used for generating random structures are shown in Table S1. In the case where a range is shown, a random value within that range is selected for building each structure. No constraints are applied during the subsequent geometry optimisation. We note that the search is not sensitive to the exact values used, given that they are physically sensible. Initial values may be inferred from typical bond lengths in existing materials or by performing a small-scale bootstrapping random search and extract typical values from low energy structures. For brevity the pairs involving P, S, and C are omitted. Full inputs for structure generations are available in the data repository at http://doi.org/10.5281/zenodo.5572347.

*Table S 2 Specie-wise minimum separations for generating random structures.*

| Species | LiFePO$_4$ (Å) | LiFeSO$_4$F (Å) | Li$_2$C$_2$O4F$_2$(Å) |
|---|---|---|---|
| Li-Li | 3.04 | 3.5-4.0 | 2.78 |
| Li-O | 2.11 | 1.90-2.10 | 1.93 |
| Li-Fe | 3.31 | 2.90-3.10 | 3.31 |
| O-O | 2.47 | 2.20-2.40 | 2.27 |
| O-Fe | 2.10 | 2.00-2.50 | 2.21 |
| Fe-Fe | 3.92 | 3.60-3.80 | 4.81 |
| Li-F | N/A | 1.8-2.1 | 1.81 |
| Fe-F | N/A | 1.8-2.2 | 2.02 |
| O-F | N/A | 2.5-2.9 | 2.82 |

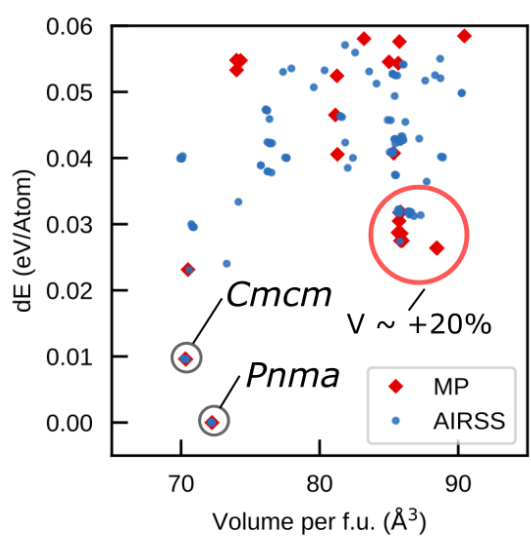

*Figure S 1Relative energy and volume of different LiFePO$_4$ phases using the PBEsol exchange-correlation functional. The cluster of less-dense low energy structures is now 25 meV/atom higher in energy compared to the olivine Pnma phase. In contrast, using PBE functional, it is only a few meV/atom higher than the Pnma phase.*

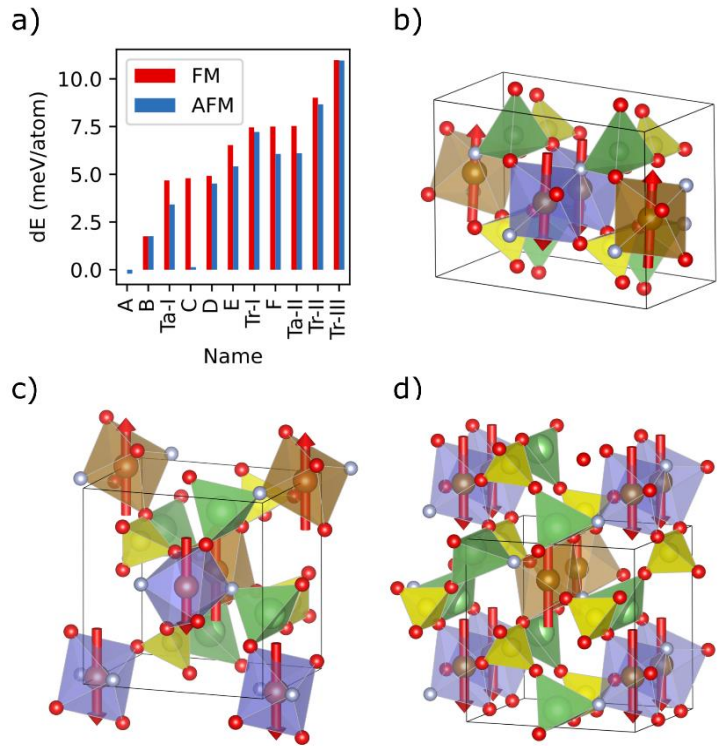

*Figure S 2 (a) Relative energy between FM and AFM spin configurations for a range of low energy polymorphs. The reference is phase A in the FM configuration. The PBE exchange-correlation functional is used. Spin polarisations on the Fe atoms are indicated by the arrows. Octahedra of spin down Fe atoms are coloured in blue. (b) The lowest energy spin confguration for Phase A.; (c) AFM spin configuration in for phase C; (d) the AFM spin configuration for phase D.*

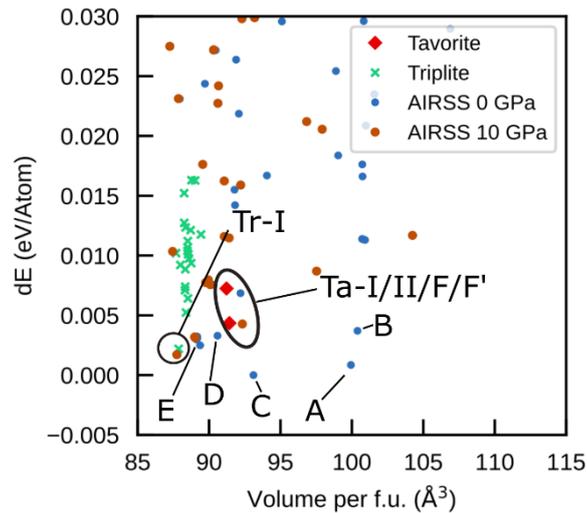

*Figure S 3 Relative energy versus volume plot for LiFeSO$_4$F polymorphs using PBEsol functional. Here the lowest energy structure is the sillimanite-II structured phase C.*

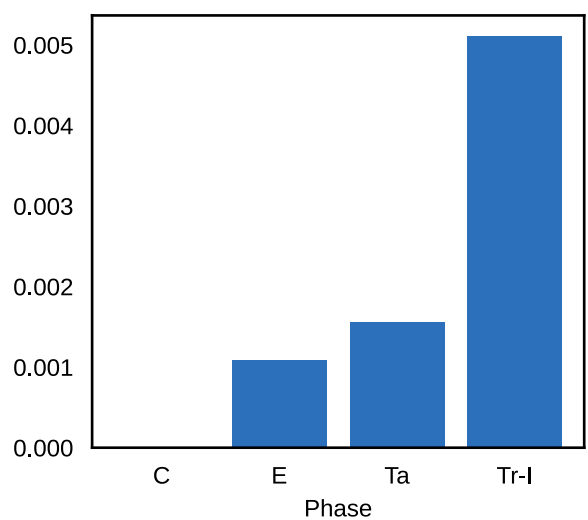

Figure S 4 Relative energies of selected structures in AFM ordering fully relaxed using the HSE06 hybrid functional.

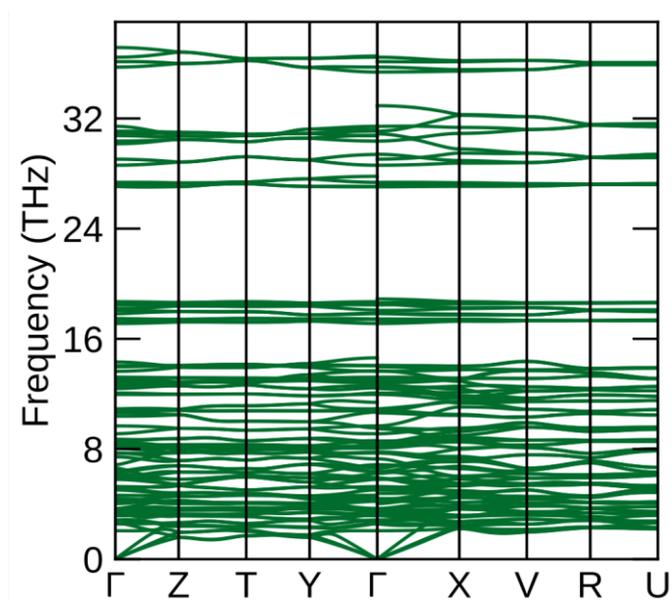

Figure S 5 Phonon band structure of LiFeSO$_4$F phase C with the AFM spin arrangement. The lack of imaginary modes suggests that this phase is dynamically stable.

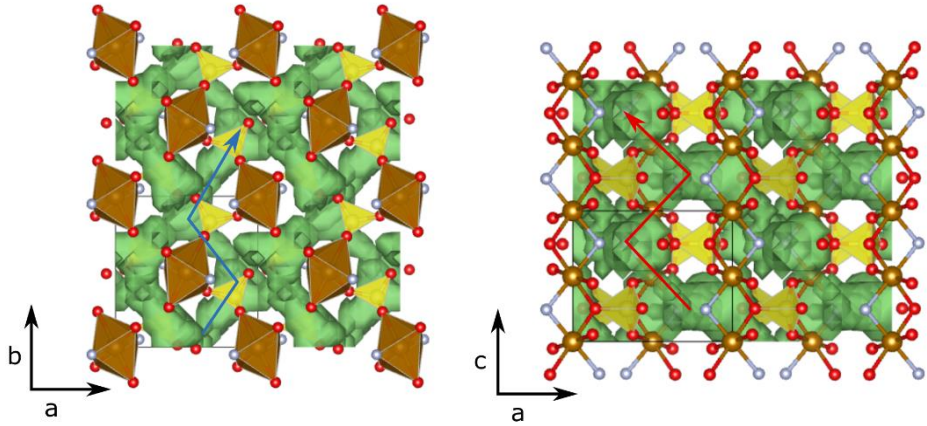

*Figure S 6 Isosurface plot of Li density from ab initio molecular dynamics simulations showing the 3D networks of the diffusion pathway.*

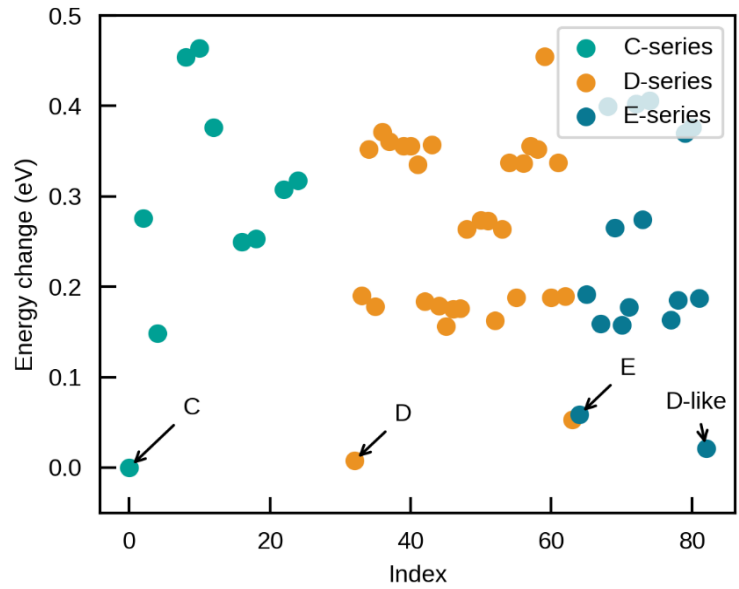

*Figure S 7 Change in energy after enumerating unique Li-Fe substitution for phase C, D, and E in the primitive cell. The minimum Li-Fe anti-site defect energy for C is in the order 0.15 eV. In addition, completely reversing the Li-Fe sites in structure E makes it into structure D and vice versa. Note that the end structures are still different in F arrangements across the $FeO_4F_2$ chains.*

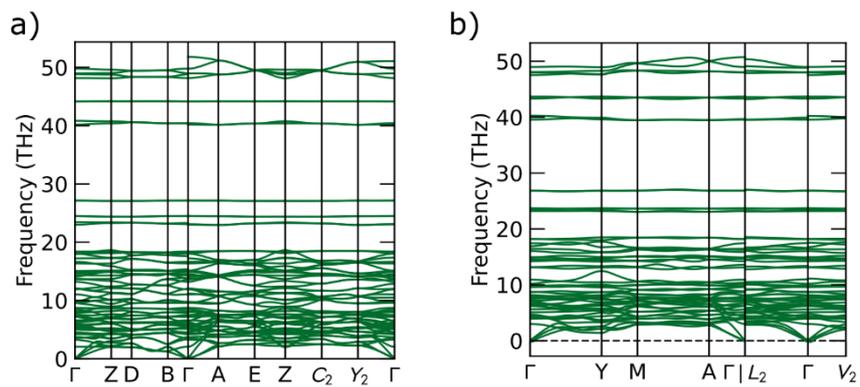

Figure S 8 The phonon band structures of (a) $Li_2FeC_2O_4F_2$ and (b) $LiFeC_2O_4F$. The lack of any imaginary modes confirms that both structures are dynamically stable.

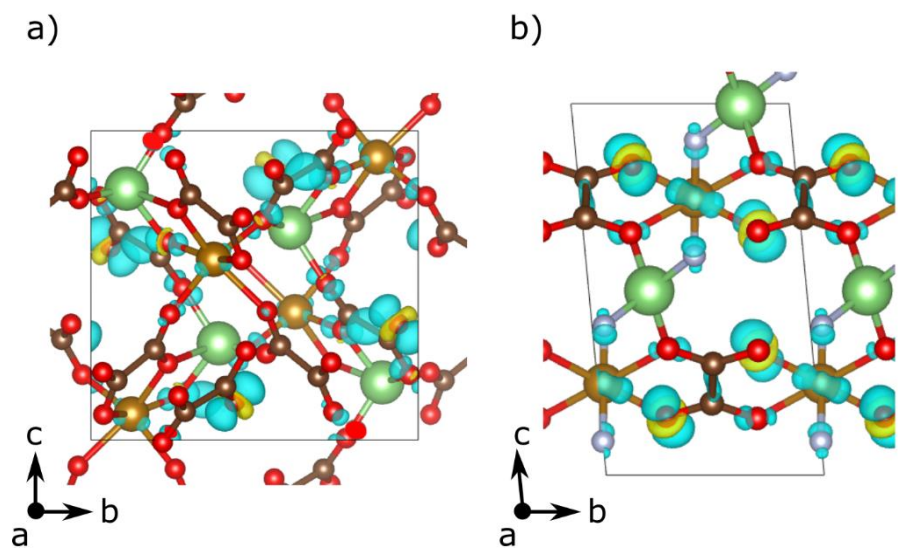

Figure S 9 The Fukui functions computed using HSE06 for (a) $Li_{2-x}Fe(C2O4)2$ and (b) $Li_{2-x}FeC2O4F2$ with x=1. The isosurface level is set to 0.008 $e^-Bohr^{-3}$ per $e^-$ removed per formula unit.

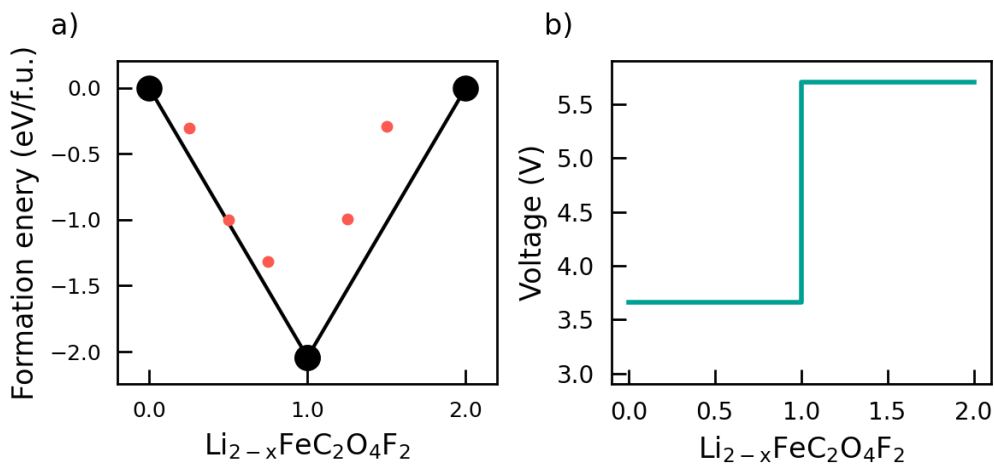

*Figure S 10 Calculated convex hull of the delithiated structures of Li$_{2-x}$FeC$_2$O$_4$F$_2$ (a) and the voltage profile constructed from it (b) using the HSE06 hybrid functional.*

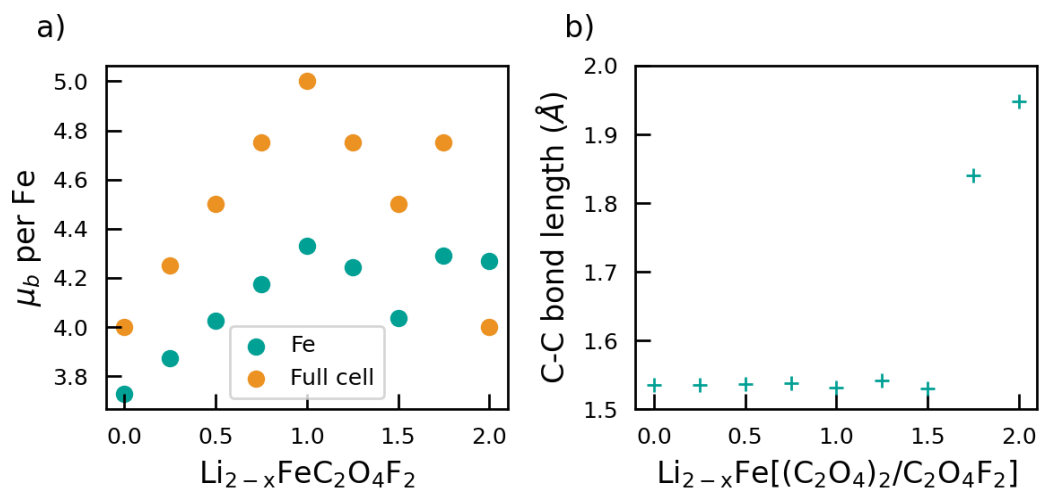

*Figure S 11 (a) Total magnetisation per Fe ("Full cell") and locally projected magnetisation (Fe) at different delithiation levels for Li$_{2-x}$FeC$_2$O$_4$F$_2$. (b) Bond lengths in delithiated structures of Li$_{2-x}$FeC$_2$O$_4$F$_2$. These results are computed using the HSE06 hybrid functional. The trends in both cases are very similar to that of Figure 10, apart from that at very high (x>1.5) delithiation levels where C=C bonds in (C$_2$O$_4$)$^{2-}$ start to break*